par **Germain ROUSSEAUX**
INLN – CNRS UMR 6638
1361 route des lucioles - 06560 Valbonne
**Richard KOFMAN**
et  **Olivier MINAZZOLI**
LPMC – CNRS UMR 6622
Université de Nice – 06108 Nice cedex 2

***RÉSUMÉ***

*Nous présentons la théorie, la modélisation et la discussion d'une expérience qui permet de choisir entre la formulation locale de Riemann-Lorenz et celle non-locale de Heaviside-Hertz afin de décrire l'Électromagnétisme Classique.*

## INTRODUCTION

Cette note fait suite à série d'articles où l'un d'entre nous (G.R.) a exposé plusieurs arguments théoriques en faveur d'une reformulation dite de Riemann-Lorenz de l'Electromagnétisme Classique en termes des potentiels contrairement à la formulation généralement enseignée dite de Heaviside-Hertz qui est basée sur les champs [1-4]. Comme dans toute remise en cause d'une théorie par une autre, seule une « experimentum crucis » permet de retenir celle décrivant les faits expérimentaux. A noter que des contribution en relation avec ce sujet sont déjà parues au Bulletin de l'Union des Physiciens [5-7].

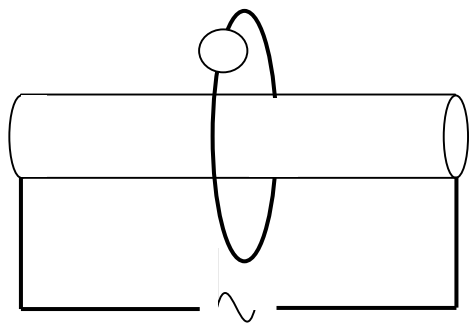

**Figure 1 :** L'expérience de Maxwell-Lodge

L'expérience qui sera l'objet de cette étude consiste à placer une spire conductrice reliée à un voltmètre autour et à l'extérieur d'un solénoïde parcouru par un courant d'intensité variable (fig.1). On observe un signal de tension dépendant du temps aux bornes du voltmètre. L'idée de l'expérience apparaît dans le Traité de James Clerck Maxwell dès 1873 [8] et la première mise en œuvre remonte à Oliver Lodge en 1889 [9]. Plusieurs reproductions [10-12] ont depuis corroboré l'existence d'un effet observable bien que les justifications théoriques aient évolué selon les auteurs [10-16]. Nous nous interrogeons sur l'origine physique de l'effet Maxwell-Lodge (M-L).

## 1. Les descriptions théoriques de l'effet Maxwell-Lodge pour un solénoïde parfait

La première description théorique est celle basée sur la formulation de Heaviside-Hertz : en partant de l'équation différentielle locale de Faraday $\partial_t \mathbf{B} = -\nabla \times \mathbf{E}$ que l'on intègre sur une surface quelconque délimitée par le circuit formé par la spire externe au solénoïde, on trouve l'équation intégrale reliant la force électromotrice (la circulation du champ électrique le long du circuit <u>fermé</u>) et le flux du champ magnétique à travers la surface [17] :

$$e = -\frac{d\Phi}{dt}$$

Le flux du champ magnétique peut être relié facilement au courant I à travers le solénoïde via le coefficient d'inductance L par la relation [17] : $\Phi = LI = \mu_0 n I \pi a^2$ où $\mu_0$ est la perméabilité du vide, n le nombre de spires par unité de longueur et a le rayon du solénoïde. Ainsi, avec une excitation sinusoïdale $I = I_0 \cos(\omega t)$ de pulsation $\omega$, on mesure expérimentalement une tension $U = e = \mu_0 n I_0 \pi a^2 \omega \sin(\omega t)$ aux bornes du voltmètre.

Malgré son étonnante simplicité, cette démonstration n'est pas satisfaisante, de notre point de vue, bien que son résultat soit corroboré par l'expérience et que toutes les formules précédentes soient valables.

Nous voulons soulever une difficulté physique, liée à la formulation de Heaviside-Hertz, qui apparaît lors de l'interprétation de cette expérience. En effet, le champ électrique **E** en un point M de la spire extérieure au solénoïde se calcule à l'aide de l'équation de Maxwell-Faraday : $\partial_t \mathbf{B}(\mathbf{r_M}) = -\nabla \times \mathbf{E}(\mathbf{r_M})$ qui est une équation <u>locale</u>. Or $\mathbf{B}(\mathbf{r_M})$ est nul pour M extérieur au solénoïde. Même si l'aspect mathématique est correct puisque la résolution de $\nabla \times \mathbf{E}(\mathbf{r_M}) = 0$ en coordonnées cylindriques permet de trouver $E_\theta = \frac{\text{Cte}}{r}$, comment un porteur de charge (un électron en l'occurrence) situé dans la spire enroulant le solénoïde et qui est soumis au champ électrique d'induction « sait-il » que le champ magnétique (qui est nul en sa position) varie à l'intérieur du solénoïde ? Plus précisément, selon l'explication précédente, un champ magnétique dans le solénoïde serait à l'origine de la création d'un champ électrique dans la spire située hors du solénoïde : si tel est le cas, ce serait un exemple d'action à distance en Electromagnétisme Classique, c'est-à-dire un effet non-local. Formulée autrement, quelle est l'origine du champ électrique local dans la spire ? La démonstration précédente montre que c'est le champ magnétique dans le solénoïde.

Cette explication est de notre point de vue inacceptable car l'Electromagnétisme est jusqu'à preuve du contraire une théorie locale. Comment peut-on, dans un phénomène temporel et en l'absence of charges statiques, rendre compte de l'apparition d'un champ électrique en un point de l'espace où le champ magnétique est absent ?

La réponse se trouve dans le Traité de Maxwell [8]: *« The conception of such a quantity, on the changes of which, and not on its absolute magnitude, the induction currents depends, occurred to Faraday at an early stage of his researches. He observed that the secondary circuit, when at rest in an electromagnetic field which remains of constant intensity, does not show any electrical effect, whereas, if the same state of the field had been suddenly produced, there would have been a current. Again, if the primary circuit is removed from the field, or the magnetic forces abolished, there is a current of the opposite kind. He therefore recognised in the secondary circuit, when in the electromagnetic field, a "peculiar electrical condition of matter" to which he gave the name of Electrotonic State.»*

Selon William Whewell (un des professeurs de Maxwell), l'état électro-tonique de Faraday se traduit par l'existence d'une quantité de mouvement dans le milieu. La résistance du milieu à la formation d'un courant est analogue à l'inertie qui s'oppose à la mise en mouvement d'un objet matériel [18] ;

$\mathbf{F} = \dfrac{d\mathbf{p}}{dt}$ ⇔ Capacité à produire un courant = variation temporelle de l'état électro-tonique

Inductance d'un circuit ⇔ Inertie d'une masse

Maxwell identifie le potentiel vecteur moderne comme étant l'intensité électro-tonique qu'il appelle aussi la quantité de mouvement électrocinétique ou électromagnétique et dont la dérivée temporelle, en l'occurrence le champ électrique, produit une force électro-motrice [8] :

$$\mathbf{F} = \dfrac{d\mathbf{p}}{dt} \Leftrightarrow \mathbf{E} = -\dfrac{d\mathbf{A}}{dt}$$

La loi empirique de Lenz (1834) sur l'induction explique le signe négatif.

Selon Maxwell : *« The Electrokinetic momentum at a point represents in direction and magnitude the time-integral of the electromotive intensity which a particle placed at this point would experience if the currents were suddenly stopped. Let Ax, Ay, Az represent the components of the electromagnetic momentum at any point of the field, due to any system of magnets or currents. Then Ax is the total impulse of the electromotive force in the direction of x that would be generated by the removal of these magnets or currents from the field, that is, if $E_x$ be the electromotive force at any instant during the removal of the system :* $A_x = \int E_x dt$ *.*

*Hence the part of the electromotive force which depends on the motion of magnets or currents in the field, or their alteration of intensity, is :* $E_x = -\dfrac{\partial A_x}{\partial t}$ $E_y = -\dfrac{\partial A_y}{\partial t}$ $E_z = -\dfrac{\partial A_z}{\partial t}$ *.*

*If there is no motion or change of strength of currents or magnets in the field, the electromotive force is entirely due to variation of electric potential, and we shall have :*

$E_x = -\dfrac{\partial V}{\partial x}$ $E_y = -\dfrac{\partial V}{\partial y}$ $E_z = -\dfrac{\partial V}{\partial z}$ *.»*

Le potentiel vecteur est une impulsion électromagnétique c.à.d. une variation de quantité de mouvement électromagnétique selon la définition mécanique d'une impulsion. Or, il est nécessaire de se donner un référentiel pour définir une impulsion mécanique ce qui se traduit par l'existence d'une constante de référence pour le potentiel vecteur. D'une manière moderne, on peut donc donner la définition suivante : le potentiel vecteur en un point M est l'impulsion qu'un opérateur extérieur doit fournir mécaniquement à une charge unité pour l'amener de l'infini, où par convention celui-ci est nul, jusqu'au point M.

Toujours selon Maxwell : *« We have now obtained in the electrotonic intensity the means of avoiding the consideration of the quantity of magnetic induction which passes through*

*the circuit. Instead of this artificial method we have the natural one of considering the current with reference to quantities exisiting in the same space as the current itself.* » En effet, le flux du champ magnétique n'est autre que la circulation du potentiel vecteur d'après le théorème de Stokes et il est tout à fait envisageable que le circuit soit dans une région exempte de champ bien qu'il entoure une zone intérieure avec champ.

Retournons à l'effet Maxwell-Lodge dans la formulation de Riemann-Lorenz. Hors du solénoïde, le champ magnétique est nul mais le potentiel vecteur est égal à un gradient [1]:

$$\mathbf{A} = \mathbf{A}_{//} = \nabla\left(\frac{\Phi\theta}{2\pi}\right) = \frac{\Phi}{2\pi r}\mathbf{e}_\theta$$

Même si le courant et le flux dans le solénoïde varient, ce potentiel vecteur non-nul ne crée pas de champ magnétique à l'extérieur du solénoïde (on se place dans l'A.R.Q.S. où il n'y a pas de champ de radiation) : $\mathbf{B} = \nabla \times \mathbf{A} = 0$, mais il crée un champ électrique :

$$\mathbf{E} = -\frac{\partial \mathbf{A}}{\partial t} = -\nabla(\frac{d\Phi}{dt}\frac{\theta}{2\pi}) = -\frac{1}{2\pi r}\frac{d\Phi}{dt}\mathbf{e}_\theta$$

qui est irrotationnel dans la spire en dehors du solénoïde ( $\mathbf{E} = \mathbf{E}_{//}$ ) : $\nabla \times \mathbf{E} = 0$ (même si le champ magnétique dans le solénoïde varie). La force électromotrice résultante (circulation du champ électrique sur le circuit de la spire) est bien indépendante du rayon de la spire ; $e = \mu_0 n I_0 \pi a^2 \omega \sin(\omega t)$. On mesure donc une tension aux bornes du voltmètre et pas une différence de potentiel scalaire dans ce cas précis.

On retrouve le résultat de la formulation de Heaviside-Hertz avec la formulation de Riemann-Lorenz à la différence cruciale que le champ électrique donc la tension mesurée dans la spire a pour origine le champ magnétique à l'intérieur du solénoïde via la médiation locale par le potentiel vecteur qui seul, est non-nul à l'extérieur du solénoïde contrairement au champ magnétique. L'intervention du potentiel vecteur est de notre point de vue essentielle à la compréhension physique du phénomène d'autant plus que le courant dans la spire résulte du transfert de quantité de mouvement du champ vers les porteurs de charge grâce au potentiel vecteur qui est bien une impulsion dans le champ [19].

## 2. Liens entre l'effet Maxwell-Lodge et l'effet Aharonov-Bohm (A-B)

Une source électronique crée des figures d'interférence dans l'expérience des fentes d'Young (dualité onde-corpuscule). Si, derrière et entre les fentes, on place un solénoïde parcouru par un courant stationnaire, celui-ci produit un champ magnétique non nul dans le solénoïde. La présence du solénoïde se traduit par une translation de la figure d'interférence due à un déphasage proportionnel au flux du champ magnétique dans le solénoïde : c'est l'effet Aharonov-Bohm vectoriel [20].

Deux écoles de pensée interprètent différemment les faits précédents selon que l'effet est attribué au champ magnétique à l'intérieur du solénoïde ou au potentiel vecteur en dehors du solénoïde [21]. Les deux écoles font l'hypothèse que le champ magnétique est nul à l'extérieur du solénoïde comme corroboré par les expériences de Tonomura et al. [22] qui ont écarté le fait que l'effet A-B soit dû à un champ magnétique de fuite ce qui avait été un contre-argument à la réalisation expérimentale de Chambers [22] où des fuites étaient présentes.

- Soit l'effet A-B est un effet non-local tel que l'onde électronique « sent » la présence du champ magnétique même si l'on s'assure par un blindage que les électrons ne pénètrent pas dans le solénoïde. Pour justifier cette interprétation, on invoque le fait que la Physique Quantique est une théorie non-locale en particulier depuis les expériences d'A.

Aspect [23]. Par ailleurs, il est dit que l' « observable » (le décalage) s'exprime uniquement en fonction du champ uniforme à l'intérieur du solénoïde et pas en fonction du potentiel vecteur non-nul à l'extérieur.

- Soit l'effet A-B est un effet local tel que l'onde électronique subit un déphasage, à l'extérieur du solénoïde, région de l'espace où le potentiel vecteur est non-nul. Ceci correspond à la formulation théorique initiale avancée par Aharonov et Bohm. L'effet A-B a ceci d'extraordinaire que l'on peut attribuer un phénomène physique au potentiel vecteur en l'absence de champ magnétique et électrique dans le cadre de la Physique Quantique. Feynman a exprimé ce point de vue sur l'effet Aharonov-Bohm dans une jonction Josephson qu'il appelle effet Mercereau [24-25] : *« What ? Do you mean to tell me that I can tell you how much magnetic field there is inside of here by measuring currents through here and here - through wires which are entirely outside - through wires in which there is no magnetic field... In quantum mechanical interference experiments there can be situations in which classicaly there would be no expected influence whatever. But nevertheless there is an influence. Is it action at distance? No, A is as real as B - realer, whatever that means. »*

Feynman utilise l'adjectif « réel » pour qualifier la nature physique du potentiel vecteur [20]. En effet, pour Feynman un champ réel est défini par *« un ensemble de nombres que l'on spécifie de telle sorte que ce qui arrive en un point dépende uniquement des nombres en ce point. On n'a pas besoin d'en savoir plus sur ce qui se passe en d'autres endroits ».* De la même manière, la circulation du champ électrique dans les phénomènes d'induction dépend de la variation temporelle du potentiel vecteur localement tout au long du circuit. L'effet M-L serait tel que l'on peut attribuer un phénomène physique (apparition d'un champ électrique local) à la présence d'un potentiel vecteur local bien que le champ magnétique local soit absent.

## 3. Discussion sur les aspects géométriques et énergétiques de l'effet Maxwell-Lodge

L'explication de l'effet M-L en terme du potentiel vecteur soulève un paradoxe quant au processus géométrique sous-jacent à l'apparition d'une force électromotrice dans un circuit. Auparavant, nous devons faire un petit détour historique. En effet, rappelons que Faraday décrivait les phénomènes d'induction d'une manière géométrique [8-27] : *« the phenomena of electromagnetic force and induction in a circuit depend on the variation of the number of lines of magnetic induction which pass through the circuit »* sachant que, selon Maxwell, *« the number of these lines is expressed mathematically by the surface-integral of the magnetic induction through any surface bounded by the circuit ».*
Poynting précisa ce point pour aider à visualiser géométriquement le potentiel vecteur [28] :
*« The assumption that if we take any closed curve, the number of tubes of magnetic induction passing through it is equal to the excess of the number which have moved in over the number which have moved out through the boundary since the beginning of the formation of the field, suggests a historical mode of describing the state of the field at any moment…One can define $A_x$, $A_y$, $A_z$ as the number of tubes of magnetic induction which would cut the axes [(Ox, Oy, Oz)] per unit length if the system were to be allowed to return to its original unmagnetic condition, the tubes now moving in the opposite direction » .*
Cette vision géométrique des phénomènes d'induction utilise la notion de lignes et de tubes. Ces quantités, qui ne sont qu'une aide à la visualisation des champs, peuvent-elles expliquer la naissance d'une fem dans la spire ? En effet, comment le flux du champ magnétique à travers la surface délimitée par la spire pourrait-il varier sans que les tubes de champ ne coupent le circuit (et donc qu'un champ magnétique soit présent en chaque point de la spire [9-28] ) ?

En effet, selon Poynting : *« Change in the total quantity of magnetic induction passing through a closed curve should always be produced by the passage of induction tubes through the curves inwards or outwards…[However], when a part of a circuit is between the poles of an electromagnet whose magnetising current is changing, we have no direct experimental evidence of the movement of induction in or out. But the induction tubes are closed, and to make them thread a circuit we might expect that they would have to cut through the boundary. The alternative seems to be that they should grow or diminish from within, the change in intensity being propagated <u>along</u> the tubes. This would be inconsistent with their closed nature [$\nabla.\mathbf{B} = 0$], unless the energy were instantaneously propagated along the whole length, and his further negatived by the theory of the transfer of energy, which implies that the energy flows transversely to the direction of the tubes. I shall suppose, then, that alteration in the quantity of magnetic induction through a closed curve is always produced by motion of induction tubes inwards or outwards through the bounding curve »*.

En plus de la difficulté géométrique, Poynting souligne une difficulté d'ordre énergétique. En effet, le vecteur de Poynting doit être dirigé transversalement à la direction des lignes de champ magnétique ce qui est une indication supplémentaire, selon lui, de la nécessité d'un mouvement des tubes à travers le circuit afin d'expliquer l'origine du transfert énergétique.

Or, précisément, la géométrie du solénoïde est telle que le flux varie non pas à cause de tubes coupant le circuit mais en étant modifié de l'intérieur du solénoïde (sans interagir avec la spire) par une variation d'intensité du courant donc du module du champ magnétique (à surface constante) qui se propage instantanément le long des tubes car le solénoïde est considéré dans l'ARQS où les effets de retard sont négligeables. Le nombre de tubes ne varie pas. De plus, le caractère solénoïdal du champ magnétique signifie que les tubes soit se referment sur eux en formant des boucles soit partent à l'infini. Dans le cadre d'un solénoïde parfait ou torique, les tubes présents à l'intérieur du solénoïde ne coupent pas la spire. Pour un solénoïde de taille finie, certes les tubes de champs se rebouclent à l'extérieur mais une variation de courant dans le solénoïde ne les fait pas se déplacer et couper la spire.

Concernant l'aspect énergétique, lorsque la spire est en circuit ouvert, le solénoïde ne transmet pas d'énergie via un vecteur de Poynting transverse étant donné qu'aucun courant ne circule dans la spire (malgré le fait que l'on détecte une tension périodique). Lorsque le circuit est fermé, il y a bien un flux d'énergie transverse via le vecteur de Poynting mais celui-ci n'est pas construit avec un hypothétique champ magnétique issu du solénoïde mais avec le champ magnétique induit dans la spire par le champ électrique crée par le potentiel vecteur. Le bilan a été examiné judicieusement par Gough & Richards [15].

Ainsi, aucun des arguments de Poynting ne tient et la vision géométrique de Faraday-Maxwell, même si elle se révèle fructueuse dans la plupart des cas, ne s'applique pas à l'effet Maxwell-Lodge. On lira avec grand intérêt l'analyse limpide et profonde de J. Roche sur cette limitation de l'explication des phénomènes d'induction en termes de lignes de champ [29].

On peut établir une analogie entre un solénoïde et un tourbillon afin de préciser notre discussion [16]. En effet, un tourbillon de vidange en mécanique des fluides peut être modélisé par les équations suivantes :  $\nabla.\mathbf{u} = 0$ équation de continuité hydrodynamique; $\mathbf{u} = \dfrac{\Gamma}{2\pi r} \mathbf{e}_\theta$ champ de vitesse en dehors du tourbillon avec $\Gamma$ la circulation de la vitesse ou le flux de vorticité; $\mathbf{u} = \dfrac{\mathbf{w}}{2} \times \mathbf{r}$ champ de vitesse dans le tourbillon avec $\mathbf{w} = \nabla \times \mathbf{u}$ la vorticité et $\mathbf{a}_{cc} = \partial_t \mathbf{u}$ l'accélération résultante due à une variation de vitesse. Pour le solénoïde en électromagnétisme, on a : $\nabla . \mathbf{A} = 0$ équation de continuité électromagnétique; $\mathbf{A} = \dfrac{\Phi}{2\pi r} \mathbf{e}_\theta$ potentiel vecteur en dehors du solénoïde avec $\Phi$ le flux de champ magnétique ou la circulation du potentiel vecteur; $\mathbf{A} = \dfrac{\mathbf{B}}{2} \times \mathbf{r}$ potentiel vecteur dans le solénoïde avec $\mathbf{B} = \nabla \times \mathbf{A}$ le champ magnétique et $\mathbf{E} = -\partial_t \mathbf{A}$ le champ électrique résultant dû à une variation de potentiel vecteur.

Admettre que le flux de champ magnétique change par une variation du nombre de tubes magnétiques impliquerait de même pour la vorticité associée au tourbillon. Or, on ne voit pas comment de la vorticité venant de l'infini pourrait se propager instantanément à travers la zone irrotationnelle pour modifier la vorticité au cœur du tourbillon et inversement. Par ailleurs, pour un tourbillon de vidange de baignoire par exemple, on peut modifier le débit de sortie en aspirant ce qui a pour conséquence l'augmentation de vorticité de l'intérieur de la zone rotationnelle (qui se traduit par une augmentation de vitesse d'où une accélération dans la zone irrotationnelle à l'extérieur du tourbillon). L'augmentation de débit est analogue à l'augmentation d'intensité de courant électrique ce qui a pour conséquence l'augmentation de champ magnétique de l'intérieur du solénoïde dans la zone rotationnelle (qui se traduit par une augmentation de potentiel vecteur donc de champ électrique dans la zone irrotationnelle à l'extérieur du solénoïde).

## 4. La modélisation d'un solénoïde de taille finie

Tout d'abord, faisons une liste des hypothèses utilisées dans le calcul de l'effet Maxwell-Lodge rappelé au §1:

- le solénoïde est infini : il n'y a pas de fuites en haut et en bas.
- le solénoïde est constitué de spires non-inclinées : il n'y a pas de fuites sur les bords.
- Les radiations (ondes) sont négligeables (donc les effets de retard) car le solénoïde fonctionne dans l'approximation des régimes quasi-stationnaire (A.R.Q.S.) : $a\omega \ll c$ où c est la célérité de la lumière.
- Les équations qui décrivent le solénoïde sont celles de la limite magnétique de Lévy-Leblond & Le Bellac [3] car la densité volumique de charge est nulle et la densité volumique de courant est non-nulle ($\rho c \ll J$).

Nous allons vérifier si les fuites d'origine géométrique (taille finie et inclinaison des spires du solénoïde) peuvent être à l'origine de la tension mesurée, ce qui serait un contre-argument à une interprétation basée sur le potentiel vecteur...

D'une manière générale, on peut identifier quatre contributions au champ magnétique total $\mathbf{B}_T$ :

$$\mathbf{B}_T = \mathbf{B}_{ARQS} + \mathbf{B}_{rad} + \mathbf{B}_f^l + \mathbf{B}_f^\alpha$$

où $\mathbf{B}_{ARQS}$ est le champ quasi-statique, $\mathbf{B}_{rad}$ le champ radiatif (tenant éventuellement compte des fuites), $\mathbf{B}_f^l$ le champ de fuite quasi-statique associé à la longueur l finie du solénoïde et $\mathbf{B}_f^\alpha$ le champ de fuite quasi-statique associé à l'inclinaison des spires d'un angle α.

Un raisonnement en ordre de grandeur permet d'évaluer l'importance relative du champ radiatif par rapport au champ quasi-statique :

$$\frac{\mathbf{B}_{rad}}{\mathbf{B}_{ARQS}} \approx \frac{a\omega}{c}$$

où dans l'approximation des régimes quasi-stationnaires, on a :

$$\frac{a\omega}{c} \approx \frac{10^{-1} \cdot 10^2}{10^8} \approx 10^{-7} \ll 1$$

On se placera donc dans l'A.R.Q.S. sachant que les champs électriques créés par induction seront reliés aux champs magnétiques par une relation du type : $\mathbf{E}^i \approx a\omega \mathbf{B}^i$.

Pour tenir compte de la géométrie du solénoïde, nous avons modélisé celui-ci par la superposition sur une longueur l de spires inclinées d'un angle γ. L'avantage de cette méthode est qu'elle permet d'accéder au champ magnétique créé à l'extérieur par la taille finie du solénoïde ; elle permet donc de rendre compte du champ de fuite dû à l'inclinaison des spires et à leur écartement. En fait, on considère des spires infiniment fines ce qui majore, dans la simulation, le champ de fuite dû à l'écartement des spires ; si ce champ de fuite majoré est négligeable, le champ de fuite vrai le sera aussi.

### 4.1. Potentiel vecteur créé par une spire de rayon a

Nous sommes dans la limite magnétique, ce qui nous permet de travailler dans la jauge de Coulomb [3]. On utilise le repère sphérique $(r,\theta,\phi)$.

Soit $J_\phi = I \sin\theta \, \delta(\cos\theta) \frac{\delta(r'-a)}{a}$, la densité volumique de courant à travers la spire.

D'après Jackson [17], le champ créé par une spire parcourue par un courant I s'écrit :

$$A_\phi = \frac{I}{sa} \int \frac{r'^2 dr' d\Omega' \sin\theta' \cos\Phi' \delta(\cos\theta') \delta(r'-a)}{\sqrt{r^2 + r'^2 - 2rr'(\cos\theta\cos\theta' + \sin\theta\sin\theta'\cos\phi')}} \quad (1)$$

avec $\frac{1}{s} = \frac{\mu_0}{4\pi}$ et $\cos\theta''\delta(\cos\theta'') = 0 \Rightarrow \sin\theta''\delta(\cos\theta'') = \delta(\cos\theta'')$ tel que :

$$A_\phi(r,\theta) = \frac{Ia}{s} \int_0^{2\pi} \frac{\cos\phi' d\phi'}{\sqrt{a^2 + r^2 - 2ar\sin\theta\cos\phi'}} \quad (2)$$

Anticipons un peu et raisonnons avec des spires pouvant être inclinées :

Les spires s'enroulant autour d'une bobine, sont inclinées de telle manière que, quelque soit $\phi$, elles semblent toujours inclinées avec le même angle (fig.2).

Soit $\gamma$ cet angle et soit $A_\phi^i$ le potentiel vecteur créé par une spire inclinée alors, par projection on obtient les composantes de A :

$$A_r = A_\phi^i \sin\gamma\cos\theta, \quad A_\theta = A_\phi^i \sin\gamma\sin\theta, \quad A_\phi = A_\phi^i \cos\gamma$$

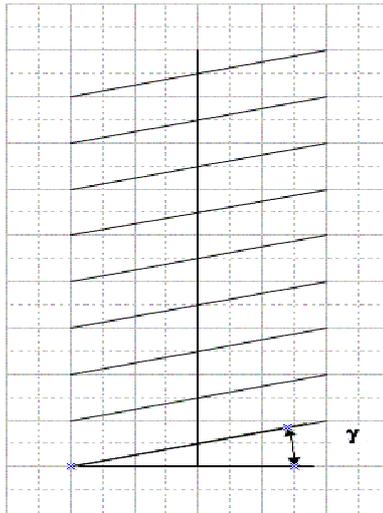
**Figure 2 :** inclinaison des spires

### 4.2. Potentiel vecteur et champs créés en un point par N spires

Pour simuler le solénoïde, il suffit maintenant de sommer, en un point M(r,θ,φ), les potentiels vecteurs créés par chaque spire. Pour des raisons de symétrie évidentes, on passe en coordonnées cylindriques (r, θ, z).

Soit L la hauteur du solénoïde, soit $\theta_i$ l'angle que fait la spire $N_i$ avec l'axe des z pour le calcul d'un point qui se trouve à une distance l du solénoïde et à une hauteur d par rapport au milieu du solénoïde (fig.3). L'écart entre chaque spire est $\varepsilon$.

Nous pouvons distinguer deux cas pour le calcul du champ créé par une spire. Le premier lorsque $\theta_i$ est inférieur à $\pi/2$, le second lorsqu'il est supérieur à $\pi/2$. Le second cas est obtenu lorsque l'on calcule le champ pour des spires supérieures à une certaine limite que nous numéroterons $N_{sup}$. On a : $N_{sup} = \mathrm{INT}\left(\dfrac{d+L/2}{\varepsilon}\right)$ ; INT étant la fonction qui renvoie la partie entière de l'argument.

Dans le premier cas, représenté sur la figure 3 par la spire, on a :

$$\theta_1 = \arctan\left(\frac{l}{d+L/2-N_1\varepsilon}\right) \text{ et } r = \frac{l}{\sin\theta_1} \qquad (5)$$

Dans le second cas :  $\theta_2 = \arctan\left(\dfrac{N_2\varepsilon-(d+L/2)}{l}\right)+\dfrac{\pi}{2}$   et   $r = \dfrac{l}{\sin\theta_2}$ \qquad (6)

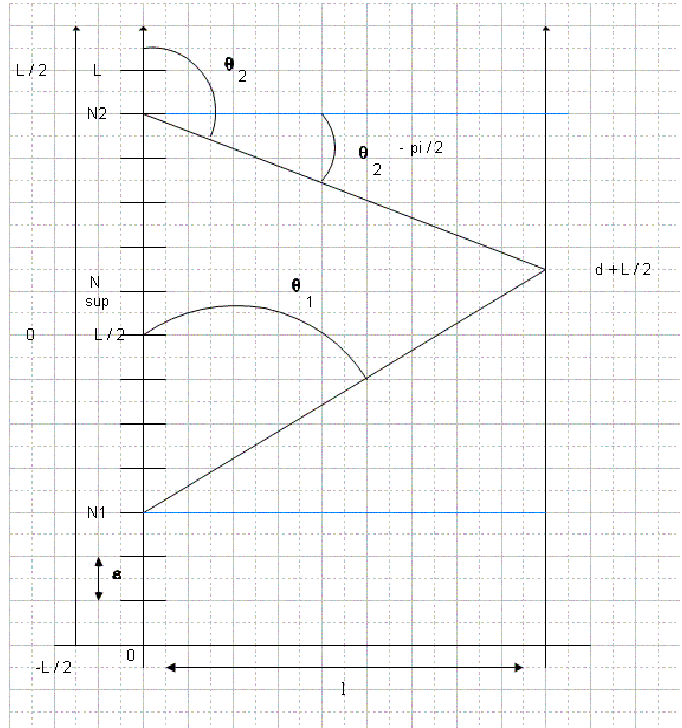
**Figure 3 :** calcul du champ créé par différentes spires en un point

Ainsi, nous pouvons calculer le potentiel vecteur créé par le solénoïde de N spires en tout point M(r, θ, z), en sommant les contributions de chaque spire, et en déduire les composantes du champ magnétique en M : $\vec{B}=\vec{\nabla}\times\vec{A}$, d'où les composantes de *B* :

$$B_r = \frac{\cos\gamma}{r\sin\theta}(\cos\theta A_\phi + \sin\theta\frac{\partial A_\phi}{\partial\theta})$$

$$B_\theta = -\frac{\cos\gamma}{r}(A_\phi + r\frac{\partial A_\phi}{\partial r}) \qquad (7)$$

$$B_\phi = \frac{\sin\gamma}{r}[(A_\phi + r\frac{\partial A_\phi}{\partial r})\sin\theta + \sin\theta A_\phi - \frac{\partial A_\phi}{\partial\theta}\cos\theta]$$

Nous allons maintenant valider nos simulations numériques en les comparant à des résultats théoriques particuliers puis nous confronterons les simulations avec les mesures effectuées au cours de l'expérience. Nous conclurons enfin sur la bonne façon, selon nous, d'interpréter l'effet Maxwell-Lodge.

## 5. Comparaison des simulations avec les prédictions théoriques

Une présentation plus complète des résultats qui suivent peut être consultée dans [30].

Comme nous pouvons le voir dans les équations définissant le potentiel vecteur (§ 1), le calcul du champ nécessite l'utilisation de méthodes d'intégration numériques. Une simple intégration par la méthode de Newton a été utilisée. Les résultats sont en très bon accord avec les expériences avec un pas spatial de 0.001. Seuls certains points nécessiteraient des méthodes informatiques plus complexes. En effet, l'axe z du solénoïde et les points se trouvant sur le bord du solénoïde sont des zones de divergence pour ces intégrales numériques.

Nous avons utilisé pour les expériences une bobine mesurant 75 cm, avec un rayon de 4.1cm. Elle possède un enroulement de 341 spires dont le fil fait 2.2 mm de diamètre. On calcule alors sa résistance R=375 $m\Omega$ ($R = \dfrac{4\phi_{solénoïde}}{\sigma \phi^2_{spire}} N$, avec N le nombre de spires du solénoïde et $1/\sigma$ la résistivité du cuivre qui vaut $1.588 \ 10^{-8} \ \Omega.m$) et son inductance L=1.08mH ($L = \mu_0 n^2 \pi \left(\dfrac{\phi_{solénoïde}}{2}\right)^2 l_{solenoide}$, avec n le nombre de spires par unité de longueur).

On détermine aussi l'angle $\gamma$ que fait une spire avec le plan xOy : $\gamma$ = 0.027 radian.

Le champ magnétique sur l'axe s'exprime selon [17]:

$B(M) = \dfrac{\mu_0 nI}{2}(\cos\alpha_1 - \cos\alpha_2)$, avec $\tan\alpha_1 = R/d$ et $\tan(\alpha_2 - \pi/2) = \dfrac{L-d}{R}$ où n est le nombre de spire par unité de longueur et I l'intensité traversant le solénoïde. Des trois composantes de B, seule $B_z$ est considérée ; $B_z$ et $B_\theta$ sont très faibles ou n'interviennent pas ($B_\theta$ sera néanmoins évoquée au paragraphe 6).

### 5.1. Composante $B_z$ en fonction de I :

L'appareillage expérimental ne nous permettant pas de mesurer le champ au milieu du solénoïde mais seulement à 10 cm (vers l'intérieur) de son extrémité, nous avons donc choisi d'effectuer le calcul théorique en ce point M (fig.4) pour une meilleure comparaison des valeurs expérimentales, théoriques et simulées.

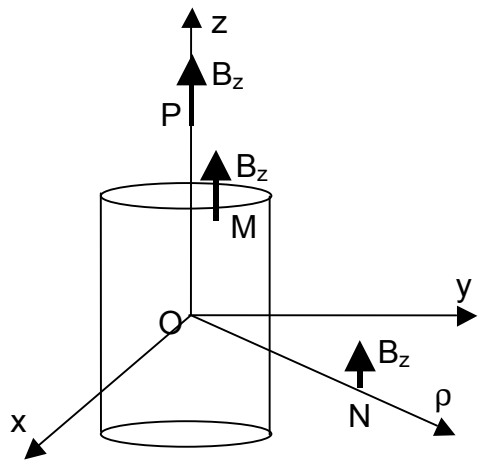

**Figure 4** : le solénoïde et $B_z$ en M (1cm, 0, 27.5cm), en N($\rho$, $\theta$, 0) sur l'axe $\rho$ et en P (0, 0, z) sur l'axe Oz.

| I (ampère) | 1 | 1,5 | 2 | 2,5 | 3 | 3,5 |
|---|---|---|---|---|---|---|
| B (gauss) | 5,5 | 8,2 | 11 | 13,7 | 16,5 | 19,2 |
| I (ampère) | 4 | 4,5 | 5 | 5,5 | 6 | 6,5 |
| B (gauss) | 22 | 24,7 | 27,4 | 30,2 | 32,9 | 35,7 |
| I (ampère) | 7 | 7,5 | 8 | 8,5 | | |
| B (gauss) | 38,4 | 41,2 | 43,9 | 46,7 | | |

**Tableau 1:** Résultats théoriques donnant $B_z$ en M, en fonction de I.

Les résultats suivants sont obtenus avec un pas de calcul de 0.001 et, pour des raisons liées à l'intégration numérique, on effectue les calculs à 1 cm de l'axe de symétrie.

| I (ampère) | 1 | 1,5 | 2 | 2,5 | 3 | 3,5 |
|---|---|---|---|---|---|---|
| B (gauss) | 5,52 | 8,27 | 11,03 | 13,79 | 16,55 | 19,31 |
| I (ampère) | 4 | 4,5 | 5 | 5,5 | 6 | 6,5 |
| B (gauss) | 22,07 | 24,82 | 27,58 | 30,34 | 33,1 | 35,86 |
| I (ampère) | 7 | 7,5 | 8 | 8,5 | | |
| B (gauss) | 38,61 | 41,37 | 44,13 | 46,89 | | |

**Tableau 2:** Résultats simulés donnant $B_z$ en M, en fonction de I avec un pas de 0.001.

L'accord entre théorie et simulation est très bon.

### 5.2. Composante $B_z$ en fonction de z :

L'allure du graphe représentant le champ magnétique sur l'axe Oz est bien connue. Nous pouvons le vérifier sur la figure 5 :

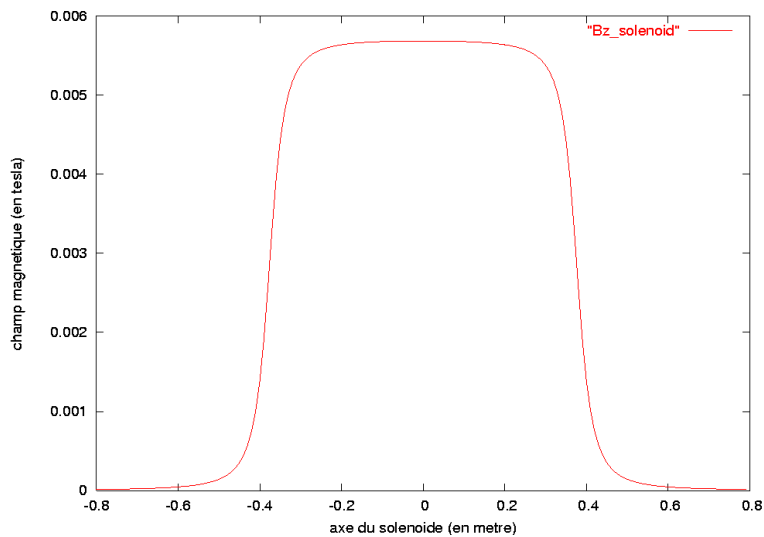

**Figure 5:** Composante $B_z$ simulée du champ magnétique

sur l'axe Oz du solénoïde (point P) pour un courant de 10A

On a une nouvelle fois un accord parfait entre la théorie et la simulation.

### 5.3. Composante $B_z$ sur le plan médian.

Le champ magnétique vaut $\mu_0 nI$ à l'intérieur du solénoïde et est proche de zéro à l'extérieur. Ceci nous donne un nouveau test pour la simulation. On a N=454.6 spires/m et I = 10 A, alors $\mu_0 nI = 57$ gauss.

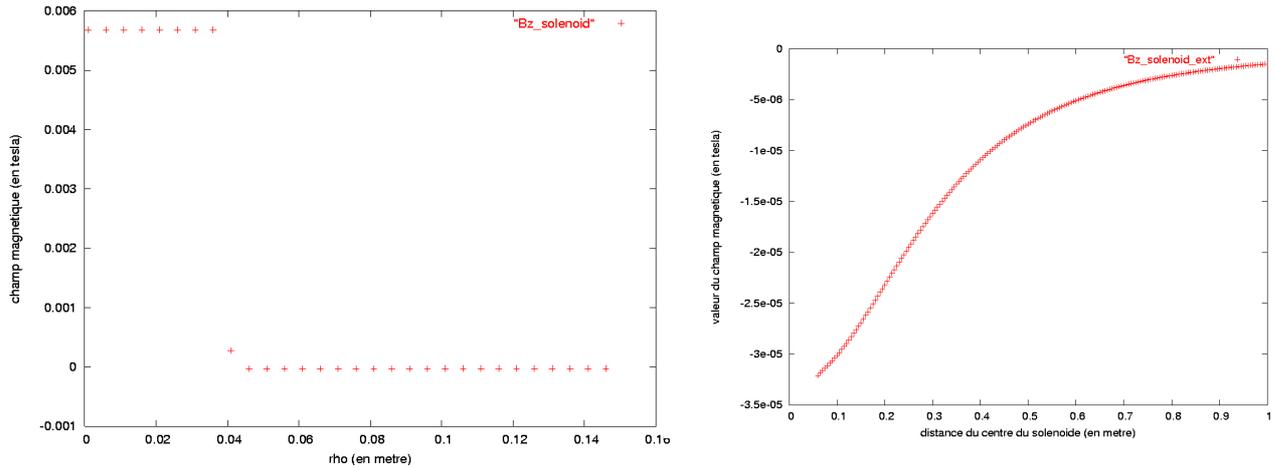

**Figure 6:** Composante $B_z$ du champ magnétique simulé sur le plan médian (point N) pour un courant de 10 A (à droite, détail x200 pour la région extérieure $\rho$ > 0.04m)

La simulation est en parfait accord avec la théorie. Un point semble cependant se distinguer, celui qui est proche de la spire à $\rho$ = 0.04 m. Cette singularité est due à un problème d'intégration numérique. Il peut être intéressant de voir comment le champ magnétique décroît avec la distance en dehors du solénoïde (ce dont nous ne nous rendons compte sur la figure 6 à cause des faibles valeurs du champ magnétique à l'extérieur comparé aux valeurs à l'intérieur) :

### 5.4. Potentiel vecteur sur le plan médian.

Soit $\vec{A}_{in}$ le potentiel vecteur à l'intérieur du solénoïde et soit $\vec{A}_{ext}$ le potentiel vecteur à l'extérieur du solénoïde. R est le rayon du solénoïde. Alors on calcule :

$$\vec{A}_{in} = \frac{\mu_0 nI\rho}{2}e_\varphi \quad \text{et} \quad \vec{A}_{ext} = \frac{\mu_0 nIR}{2}\left(\frac{R}{\rho}\right)e_\varphi \qquad (1)$$

En R, la valeur du potentiel vecteur est de 1.1 gauss.m pour un courant de 10 ampères.

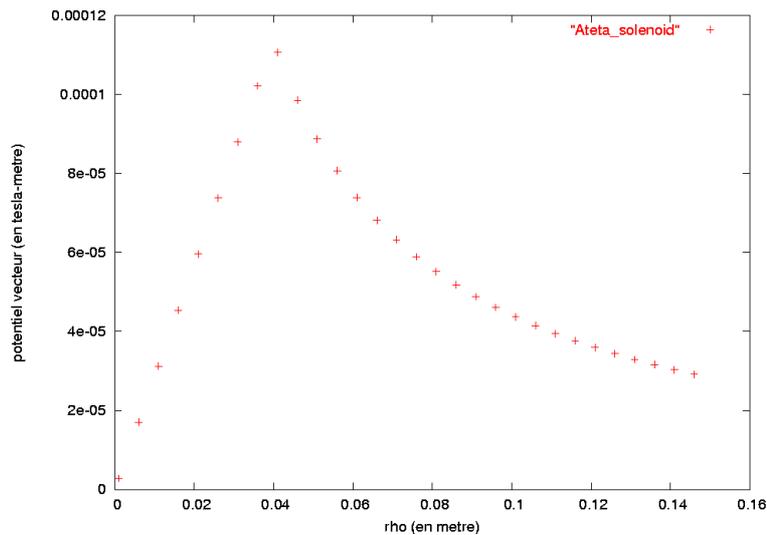

**Figure 7:** Composante $A_\theta$ du potentiel vecteur simulé sur le plan médian (point N) pour un courant de 10 A.

Nous retrouvons bien dans ce graphique (fig.7) une partie linéaire à l'intérieur du solénoïde et une partie décroissante en 1/☐ à l'extérieur. Là encore, il y a un parfait accord entre la théorie et la simulation.

## 6. Comparaison des simulations avec les résultats expérimentaux

Le champ a été mesuré à l'aide d'un gauss-mètre F.W. BELL (modèle 4048) et de deux sondes cal. n°1760 pour des mesures axiales et cal. n°1450 pour des mesures latérales.

L'incertitude sur les valeurs mesurées est de 0.2 gauss. Ce qui nous donne une incertitude relative importante lorsque l'on se place sur le plan médian puisque les valeurs du champ sont de l'ordre du dixième de gauss. Il arrive que malgré l'incertitude nous indiquions qu'il y a un champ de 0.1 gauss car, en effet, nous pouvons observer une légère variation entre les mesures avec et sans courant.

### 6.1. Mesure du champ magnétique sur l'axe

On mesure le champ à 10 cm à l'intérieur du solénoïde (point M), comme précédemment dans le § 5.

| I (Ampère) | 1 | 1,5 | 2 | 2,5 | 3 | 3,5 |
|---|---|---|---|---|---|---|
| B (gauss) | 5,5 | 8,2 | 10,8 | 13,4 | 16,2 | 18,9 |
| I (Ampère) | 4 | 4,5 | 5 | 5,5 | 6 | 6,5 |
| B (gauss) | 21,7 | 24,4 | 27,1 | 29,8 | 32,4 | 35,3 |
| I (Ampère) | 7 | 7,5 | 8 | 8,5 | | |
| B (gauss) | 37,9 | 40,7 | 43,4 | 46,1 | | |

**Tableau 3:** Résultats expérimentaux donnant $B_z$ en M

Ces résultats se comparent favorablement à ceux du tableau 2.

### 6.2. Mesure du champ magnétique à l'extérieur du solénoïde

Pour finir l'évaluation de la simulation, nous avons fait une série de mesures de $B_z$ à l'extérieur du solénoïde. Cette évaluation est d'une importance capitale puisque le but de la simulation est de pouvoir prédire la valeur du champ magnétique extérieur en tout point. La précision du gauss-mètre est à sa limite pour certains points, mais elle permet d'avoir une allure du champ à l'extérieur (fig.8).

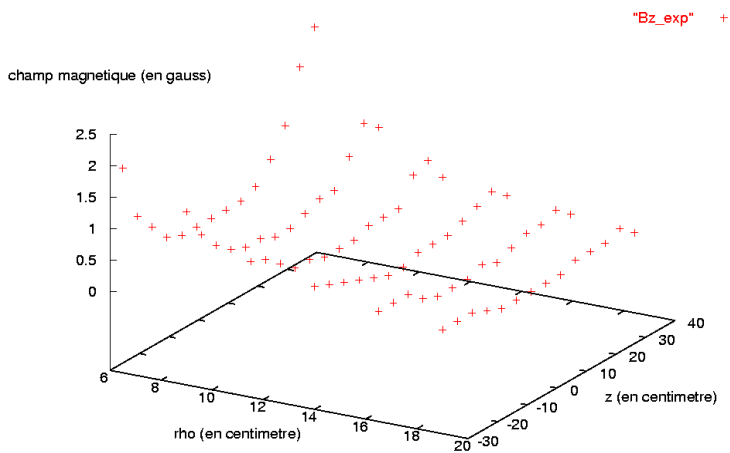

**Figure 8 :** mesure de la composante $B_z$ du champ magnétique
à l'extérieur du solénoïde pour un courant de 10 A

Les valeurs de $B_z$ simulé à l'extérieur sont données sur la figure 9 ; elles sont en bon accord avec les valeurs tirées de l'expérience (fig.8).
Les simulations sont donc d'une part en accord avec la théorie, d'autre part en accord avec l'expérience. Nous utiliserons donc dans la suite les résultats obtenus par simulation.

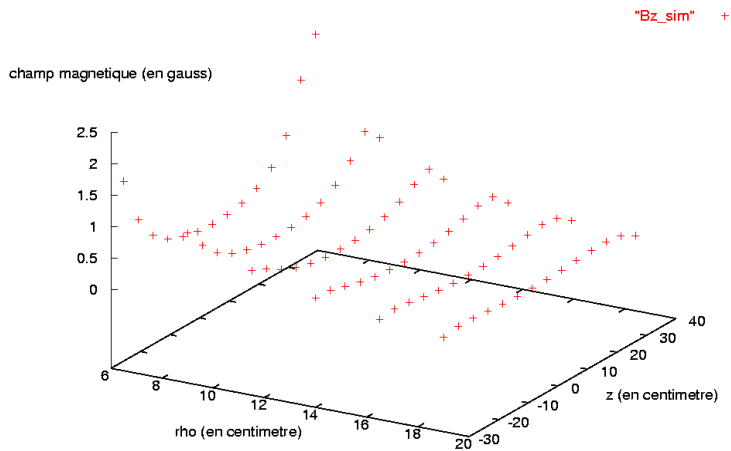

**Figure 9 :** simulation de la composante $B_z$ du champ magnétique
à l'extérieur du solénoïde pour un courant de 10 A

Les spires étant inclinées, le potentiel vecteur possède, en plus d'une composante $A_\theta$, une composante $A_z$. Ceci implique que le champ magnétique possède trois composantes comme nous pouvons le voir dans les équations (§ 1).

Ainsi, l'intérêt du modèle « solénoïde=somme de spires inclinées » est qu'il tient compte, à l'extérieur du solénoïde, de la composante $B_\theta$ du champ magnétique trouvée expérimentalement.

Ces résultats montrent que le modèle « solénoïde = somme de spires inclinés » est donc un très bon modèle car il permet d'avoir la précision nécessaire pour la discussion que nous souhaitons avoir sur la réalité du potentiel vecteur.

Une visualisation frappante du module de B à l'extérieur est donnée en exprimant le résultat des simulations en niveaux de gris (fig.10).

## 6.3. Fem induite dans une boucle conductrice extérieure au solénoïde

Le champ magnétique sur le plan médian, là où nous placerons les spires pour mesurer la f.e.m induite, est inférieur à 1% de la valeur de B à l'intérieur du solénoïde. Le champ magnétique de fuite est-il, pour autant, négligeable dans le phénomène observé ? Cela paraîtrait logique mais demande toutefois à être vérifié.

Le champ magnétique de fuite est dû à l'existence d'un potentiel vecteur de fuite puisque $\vec{B}=\vec{\nabla}\times\vec{A}$. Ce potentiel vecteur de fuite (que nous noterons $\vec{A}_f$) peut être également déduit de la simulation. En effet, on a :

$$A_f = A_{simulé} - A_{théorique} \qquad (2)$$

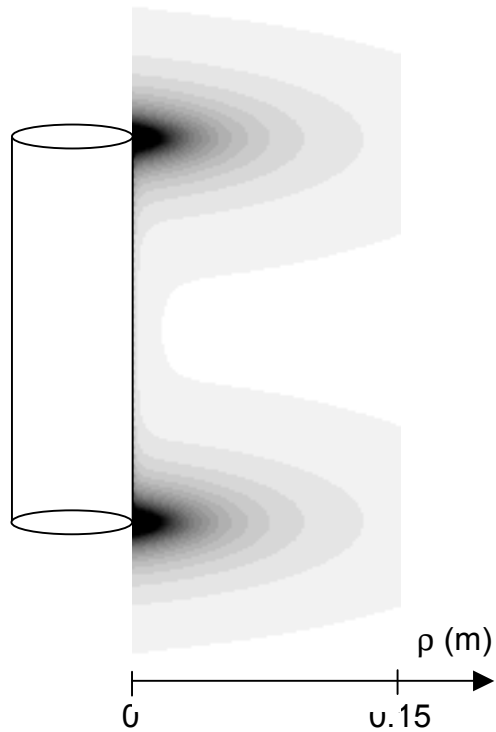

**Figure 10 :** image en niveaux de gris du module du champ magnétique à l'extérieur du solénoïde. B varie de 0.6 Gauss à chaque saut de niveau de gris (le lobe noir est exagéré afin d'améliorer la visibilité pour B faible)

Alors on en déduit le champ électrique de fuite à partir de $\vec{E}=-\partial_t\vec{A}$ :

$$E_f = \omega A_f \quad (3)$$

Et ainsi on a la force électromotrice due aux fuites avec $e=\oint \vec{E}\cdot d\vec{l}$ :

$$e_f = 2\pi\rho\omega A_f \quad (4)$$

On a, par ailleurs, la force électromotrice théorique :

$$e_{th} = 2\pi\rho\omega A_{th} \quad (5)$$

Soit, avec (1) :

$$e_{th} = \pi\omega\mu_0 n I R^2 \quad (6)$$

On remarque que la f.e.m. théorique ne dépend pas de l'éloignement au solénoïde.

| I (ampère) | f ( Hz) | fem (mV) |
|---|---|---|
| 1 | 1601 | 30,3 |
| 1,01 | 1500 | 28,7 |
| 1,01 | 1400 | 26,8 |
| 1,01 | 1300 | 24,9 |
| 1,01 | 1200 | 23 |
| 1,01 | 1099 | 21 |

**Tableau 4:** valeurs théoriques de la f.e.m induite

| I (ampère) | f ( Hz) | fem 1 (mV) | fem 2 (mV) | fem 3 (mV) | fem 4 (mV) | fem 5 (mV) |
|---|---|---|---|---|---|---|
| 1 | 1601 | 28 ± 3 | 28 ± 3 | 28 ± 3 | 27 ± 3 | 26 ± 3 |
| 1,01 | 1500 | 26 ± 2 | 26 ± 4 | 25 ± 3 | 24 ± 2 | 24 ± 2 |
| 1,01 | 1400 | 24 ± 2 | 24 ± 2 | 24 ± 2 | 24 ± 3 | 23 ± 3 |
| 1,01 | 1300 | 23 ± 2 | 22 ± 2 | 23 ± 2 | 22 ± 2 | 22 ± 2 |
| 1,01 | 1200 | 21 ± 2 | 21 ± 3 | 21 ± 2 | 20 ± 2 | 20 ± 2 |
| 1,01 | 1099 | 19 ± 2 | 19 ± 2 | 18 ± 2 | 19 ± 2 | 18 ± 3 |

**Tableau 5 :** valeurs expérimentales de la f.e.m induite

Les f.e.m 1,2,3,4,5 sont respectivement les f.e.m mesurées pour une spire de rayon 5cm, 7.5 cm, 10 cm, 12.5 cm, 15 cm.

| I (ampère) | f ( Hz) | fem 1 (mV) | fem 2 (mV) | fem 3 (mV) | fem 4 (mV) | fem 5 (mV) |
|---|---|---|---|---|---|---|
| 1 | 1601 | 30,1 | 29,8 | 29,4 | 28,8 | 28,2 |
| 1,01 | 1500 | 28,2 | 27,9 | 27,5 | 27 | 26,5 |
| 1,01 | 1400 | 26,3 | 26 | 25,7 | 25,2 | 24,7 |
| 1,01 | 1300 | 24,4 | 24,2 | 23,8 | 23,4 | 22,9 |
| 1,01 | 1200 | 22,6 | 22,3 | 22 | 21,6 | 21,2 |
| 1,01 | 1099 | 20,7 | 20,4 | 20,2 | 19,8 | 19,4 |

**Tableau 6 :** valeurs simulées de la f.e.m induite

| I (ampère) | f ( Hz) | fem 1 (mV) | fem 2 (mV) | fem 3 (mV) | fem 4 (mV) | fem 5 (mV) |
|---|---|---|---|---|---|---|
| 1 | 1601 | -0,2 | -0,5 | -1 | -1,5 | -2,1 |
| 1,01 | 1500 | -0,2 | -0,5 | -0,9 | -1,4 | -2 |
| 1,01 | 1400 | -0,2 | -0,5 | -0,8 | -1,3 | -1,8 |
| 1,01 | 1300 | -0,2 | -0,4 | -0,8 | -1,2 | -1,7 |
| 1,01 | 1200 | -0,2 | -0,4 | -0,7 | -1,1 | -1,6 |
| 1,01 | 1099 | -0,2 | -0,4 | -0,7 | -1 | -1,4 |

**Tableau 7 :** valeurs simulées de la f.e.m de fuite induite

Ainsi, le champ magnétique de fuite joue très clairement un rôle dans l'effet M–L, mais il n'explique pas l'effet en lui même (Figure 11). En effet, lorsque l'on est proche du solénoïde, le champ magnétique de fuite ne peut expliquer la f.e.m. que l'on mesure sur la spire à l'extérieur du solénoïde. Au contraire, la f.e.m. due aux fuites est en opposition de phase avec la f.e.m. théorique. De cette manière la f.e.m. totale diminue à mesure que la f.e.m de fuite augmente puisque la f.e.m théorique est constante sur le plan médian. On a alors une décroissance de la f.e.m totale, à mesure que l'on s'éloigne du solénoïde. Aussi, il n'est pas étonnant de voir la f.e.m. totale (c.a.d. celle que l'on mesure) diminuer quand le rayon de la spire augmente car on a de plus en plus de champ parasite « rebouclant » dans cette surface et annulant peu à peu les effets du champ au cœur du solénoïde. Nous avons donc une explication locale au phénomène qui tend à diminuer la f.e.m. induite dans une spire quand le rayon de celle-ci augmente.

On pense parfois que la f.e.m. que l'on mesure aux bornes de la spire extérieure est due au champ magnétique de fuite à cause de la taille non infinie du solénoïde, ou d'un effet de l'inclinaison des spires non prévu par la théorie, ou encore des deux en même temps.

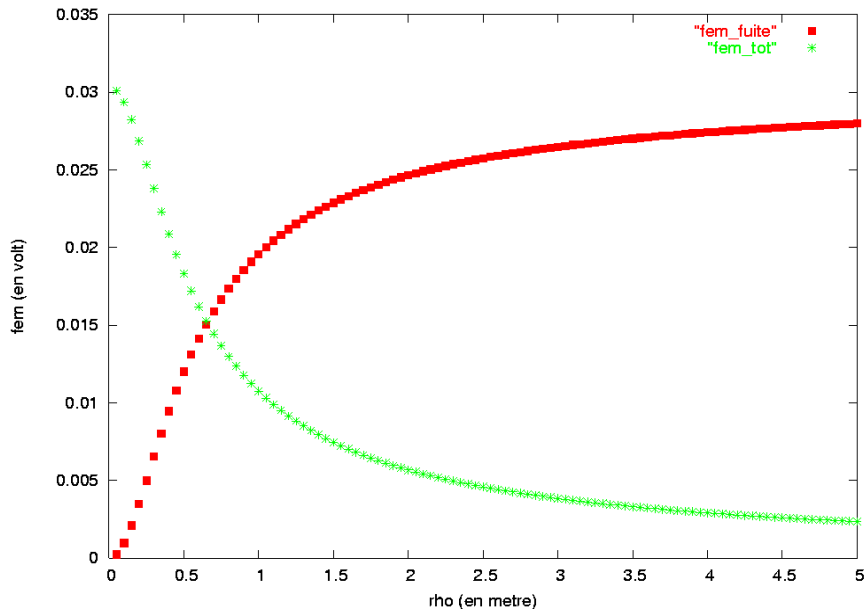

**Figure 11 :** modules simulés de la f.e.m de fuite et f.e.m totale aux bornes d'une spire extérieure au solénoïde pour un courant sinusoïdal de 1 A à 1600 Hz

Or, ce n'est pas le cas. La taille finie du solénoïde intervient en faisant décroître la f.e.m avec l'éloignement et l'inclinaison des spires a pour effet de faire apparaître une composante $B_\theta$ du champ magnétique (qui n'intervient donc pas dans le calcul de la f.e.m) et de légèrement modifier la valeur de la composante $B_z$ du champ magnétique à l'extérieur mais de manière négligeable.

## 8- Conclusion

Nous avons proposé dans cet article une description locale complète de l'effet Maxwell-Lodge d'où il ressort que l'utilisation du potentiel vecteur est nécessaire pour interpréter cet effet. Dès lors, la notion de champ « réel » proposée par Feynman [20] s'applique au potentiel vecteur qui passe du statut « d'outil mathématique » permettant le calcul des champs électrique et magnétique à celui de champ à part entière. Ainsi, la formulation de Riemann–Lorentz semble plus naturelle que la formulation d'Heaviside–Hertz puisqu'elle présuppose les équations de Riemann pour les potentiels comme équations fondamentales de l'électromagnétisme alors que la formulation de Heaviside-Hertz les considère comme équations secondaires. En conclusion, les faits expérimentaux imposent de reformuler l'électromagnétisme classique en fonction des potentiels en tant que quantités plus fondamentales que les champs qui en dérivent et non l'inverse.

**BIBLIOGRAPHIE**

[1]     ROUSSEAUX G. & GUYON E., A propos d'une analogie entre la mécanique des fluides et l'électromagnétisme, *Bulletin de l'Union des Physiciens*, 2002, 841, p. 107-136
http://www.udppc.asso.fr/bup/841/0841D107.pdf

[2]     ROUSSEAUX G., On the physical meaning of the gauge conditions of Classical Electromagnetism: the hydrodynamics analogue viewpoint, *Annales de la Fondation Louis de Broglie*, 2003, 28, p. 261-270
http://www.ensmp.fr/aflb/AFLB-282/ablb282p261.htm

[3]     ROUSSEAUX G. & A. DOMPS, Remarques supplémentaires sur l'approximation des régimes quasi-stationnaires en électromagnétisme, *Bulletin de l'Union des Physiciens*, 2004, 863
http://www.udppc.asso.fr/bup/863/08630621.pdf

[4]     ROUSSEAUX G., Sur la théorie de Riemann-Lorenz de l'Électromagnétisme Classique, *Bulletin de l'Union des Physiciens*, 2004, 868, p. 41-46

[5]     FLECKINGER R., CARLES R. et PEREZ J.P., Faut-il, en régime quasi-stationnaire, « tuer » la différence de potentiel ? *Bulletin de l'Union des Physiciens*, 1990, 722, p. 375-384

[6]     FLECKINGER R., CARLES R. et PEREZ J.P., « Tensions » sur la différence de potentiel ? *Bulletin de l'Union des Physiciens*, 1991, 732, p. 567-570

[7]     BOUSSIÉ A., Quelle tension mesure un voltmètre en régime quasistatique ? *Bulletin de l'Union des Physiciens*, 1992, 744, p. 757-770

[8]     MAXWELL J. CLERK, A treatise on electricity and magnetism, Volume I & II, Dover Publications, 1954

[9]     LODGE O., On an Electrostatic Field Produced by Varying Magnetic Induction, *Phil. Mag.*, 1889, 27, p. 469-478

[10]    ROMER R.H., What do "voltmeters" measure ?: Faraday law in a multiply connected region, *American Journal of Physics*, 1982, 50, p. 1089-1093

[11]    LANZARA E. & ZANGARA R., Potential difference measurements in the presence of a varying magnetic field, *Physics Education*, 1995, 30, p. 85-89

[12]    IENCINELLA D. & MATTEUCCI G., An introduction to the vector potential, *Eur. Jour. of Physics*, 2004, 25, p. 249-256

[13]   KONOPINSKI E.J., What the electromagnetic vector potential describes, *American Journal of Physics*, 1978, 46 (5), p. 499-502

[14]   SEMON M.D. & TAYLOR J.R., Thoughts on the magnetic vector potential, *American Journal of Physics*, 1996, 64 (11), p. 1361-1369

[15]   GOUGH W. & RICHARDS J.P.G., Electromagnetic or electromagnetic induction ?, European Journal of Physics, 1986, 7, p. 195-197

[16]   JECH B., Variations sur le potentiel vecteur IV, *Cahier Enseignement Supérieur du Bulletin de l'Union des Physiciens*, 2001, 830 (2), p. 85-101

[17]   JACKSON J.D., Classical Electrodynamics, third edition, John Wiley & Sons, Inc., 1998.

[18]   ANDERSON R., On an Early Application of the Concept of Momentum to Electromagnetic Phenomena: The Whewell-Faraday Interchange, *Studies in the History and Philosophy of Science*, 1994, 25, p. 577-594

[19]   TRAMMEL G.T., Aharonov-Bohm Paradox, *Physical Review*, 1964, 134, 5B, p. B1183-B1184

[20]   FEYNMAN R., LEIGHTON R. & SANDS M., *The Feynman Lectures on Physics*, 1964, 2, p. 15-7/15-14, Addison Wesley, Reading, Ma.

[21]   EYNCK, T.O., LYRE H. & RUMMELL N., A versus B! Topological nonseparability and the Aharonov-Bohm effect, 2001.
http://philsci-archive.pitt.edu/archive/00000404/00/Ab.pdf

[22]   TONOMURA A., The quantum world unveiled by electron waves, World Scientific, 1998.

[23]   ASPECT A. et al., Experimental Tests of Realistic Local Theories via Bell's Theorem, *Physical Review Letters*, 1981, 47, p. 460

[24]   GOODSTEIN D. & GOODSTEIN J., Richard Feynman and the History of Superconductivity, *Physics in Perspective*, 2000, 2, Issue 1, p. 30-47

[25]   JAKLEVIC R.C., LAMBE J.J., SILVER A.H. and MERCEREAU J.E., Quantum interference from a static vector potential in a field-free region, *Physical Review Letters*, 1964, 12, Number 11, p. 274-275

[26]   MAXWELL J. CLERK, On Physical Lines of Force (1861-2), W.D. Niven, ed., *The Scientific Papers of James Clerk Maxwell*, 1890, 2 vols., New York

[27]   MAXWELL J. CLERK, On a method of making a direct comparison of electrostatic with electromagnetic force; with a note on the electromagnetic theory of light, W.D. Niven, ed., *The Scientific Papers of James Clerk Maxwell*, 1890, 2 vols., New York

[28] POYNTING J.H., On the connection between electric current and the electric and magnetic inductions in the surrounding field, *Phil. Trans*., 1885, 176, p. 277-306

[29] ROCHE J., Explaining electromagnetic induction : a critical re-examination. The clinical value of history in physics, *Physics Education*, 1987, 22, p. 91-99

[30] MINAZZOLI Olivier, Champ magnétique et potentiel vecteur créés par un solénoïde, Rapport de stage effectué au LPMC (UMR 6622), juin-septembre 2004.
http://mlilom.perso.libello.com  ou  http://www.inln.cnrs.fr/~rousseaux/

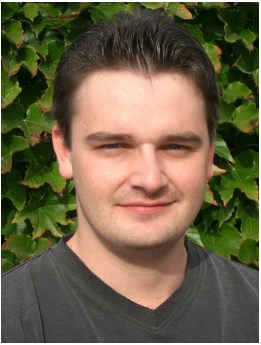

Germain Rousseaux est chercheur post-doctoral à l'Université de Nice – Sophia Antipolis; ses spécialités sont l'électromagnétisme, l'hydrodynamique et la culture scientifique.

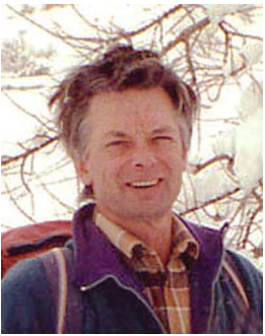

Richard Kofman est professeur de Physique à l'Université de Nice – Sophia Antipolis; ses spécialités sont l'électromagnétisme, l'optique et les nanostructures.

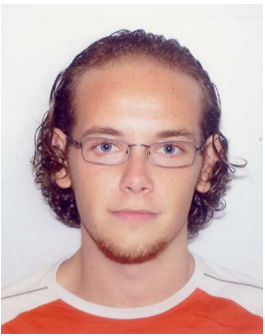

Olivier Minazzoli est étudiant en Master de Physique à l'Université de Nice – Sophia Antipolis